# MUON CAPTURE FOR THE FRONT END OF A $\mu^+$-$\mu^-$ COLLIDER*

D. Neuffer, Fermilab, Batavia, IL 60510, USA
C. Yoshikawa, Muons, Inc., Batavia IL 60510, USA

*Abstract*

We discuss the design of the muon capture front end for a $\mu^+$-$\mu^-$ Collider. In the front end, a proton bunch on a target creates secondary pions that drift into a capture transport channel, decaying into muons. A sequence of rf cavities forms the resulting muon beams into strings of bunches of differing energies, aligns the bunches to (nearly) equal central energies, and initiates ionization cooling. The muons are then cooled and accelerated to high energy into a storage ring for high-energy high luminosity collisions. Our initial design is based on the somewhat similar front end of the International Design Study (IDS) neutrino factory.

## INTRODUCTION

The goal of a $\mu^+$-$\mu^-$ collider is to provide high-energy high-luminosity collisions of μ bunches. It consists of:

- a proton source with a baseline intensity goal of 4 MW beam power (15Hz, ~10GeV protons, ~2ns bunches. (~$3\times10^{14}$p/ pulse),
- a target, capture and initial cooling section that produces π's that decay into μ's and captures them into a small number of bunches.
- a cooling section that reduces the beam emittances to collider intensities by ionization cooling.
- an accelerator that takes the μ's to high energy and inserts them into a storage ring for collisions within a high-sensitivity detector.

The present paper discusses the muon capture and initial cooling system.

The front end is based on the similar front end of a neutrino factory, as presented in the IDS design study,[1] but reoptimized toward the somewhat different requirements of a collider system. In this paper we first present and discuss the IDS design and then discuss extensions toward a $\mu^+$-$\mu^-$ collider front end. We then present a collider version and discuss future R&D directions and efforts. In this system we follow ref. [1], and set 201.25 MHz as the baseline bunch frequency. The π's (and resulting μ's) are initially produced with broad energy spreads, much larger in phase space than a 200 MHz rf bucket. In this "front end" system, we capture these μ's into strings of ~200 MHz bunches, rotate the bunches to equal energies, and cool them for acceleration to full energy. The method captures both $\mu^+$'s and $\mu^-$'s simultaneously and can be adapted to feed a $\mu^+$-$\mu^-$ collider. The major difference is that for a collider we will need to recombine the trains of $\mu^+$'s and $\mu^-$'s to single (or a few) bunches for maximal luminosity.

*Research supported by US DOE under contract DE-AC02-07CH11359 and SBIR grant DE-SC-0002739.

## ν-FACTORY FRONT END

The IDS front end is shown in Figure 1 at the end of this paper. ~10 GeV protons are targeted onto a Hg jet target that is encapsulated in a 20 T solenoid. π's created from the target are captured as they traverse the ~15m long solenoid that has a field profile that starts at 20T and 7.5cm radius at the target and tapers off to ~1.5T and 30cm radius at the end.

The taper is followed by a Drift section, where π's decay to μ's, and the bunch lengthens, developing an energy/time correlation. The following Buncher and Rotator sections contain rf cavities within focusing ~1.5T solenoids. (fig. 2) In the Buncher, rf voltages are applied to the beam to form it into a string of bunches of different energies. This is obtained by requiring that the rf wavelength of the cavities be set to an integer fraction of the cτ between reference particles:

$$\lambda_{rf}(L) = \frac{L}{N}\left(\frac{1}{\beta_N} - \frac{1}{\beta_0}\right)$$

In the IDS baseline, muons with $p_0$ = 232 MeV/c, and $p_N$ = 154 MeV/c at N=10 are used as reference particles. The rf frequency decreases from 320 to 232 MHz along the 33m Buncher while the rf gradient in cavities increases from 0 to 9 MV/m. In the Rotator, the lower energy reference particle is moved to an accelerating phase as the wavelength separation is also lengthened. (10 → ~10.05) and the rf gradient increases to 12 MV/m. At the end of the Rotator the μ's have been formed into a train of 201.25 MHz bunches with average momenta of ~232MeV/c and $\delta p_{rms}/p \approx$ 10%. The bunch train is ~80m long with ~50bunches.

The μ's are matched into a cooling section (fig. 3) which consists of rf cavities, LiH absorbers for cooling and alternating solenoids (AS) for focusing. ($B_z$ oscillates from 2.7 to -2.7T with a 1.5 m period.). The cooling doubles the number of accepted μ's while reducing the rms transverse emittances by a factor of 3. After ~75m of cooling, we find that the system accepts ~0.1$\mu^+$/ 10 GeV proton within reference acceptances of $\varepsilon_{L,N}$ <0.15m, $\varepsilon_{t,N}$ <.3cm, which are the acceptances of the downstream acceleration and storage rings. As a bonus, the method simultaneously produces bunch trains of both signs ($\mu^+$ and $\mu^-$) at equal intensities. As a result, the IDS neutrino factory is designed to use both signs in its acceleration and decay ring. The capture of both signs implies that it should be possible to extend the neutrino factory front end to initiate a collider, where simultaneous $\mu^+$ and $\mu^-$ bunches are required.

The µ capture concept requires using relatively high gradient rf fields interleaved with relatively strong solenoidal magnetic fields. Presently, we are uncertain what gradients are possible within magnetic fields and alternate configurations using "bucked-field" or magnetic insulation or gas-filled rf cavities may be needed. In the present scenario, we assume that the solution is found and that gradients can be extended from the NF solution to higher levels in the second-generation $\mu^+$-$\mu^-$ Collider front end.

This method obtains IDS bunch trains of ~80m length or ~56 200MHz bunches, but with ~67% of the µ's within the first ~20 bunches. A collider would need to recombine these bunches and that task would be eased if the bunch train were shortened.

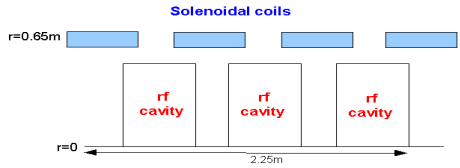

**Figure 2**: Radial projection of a Buncher and Rotator section, with rf cavities and solenoidal focusing.

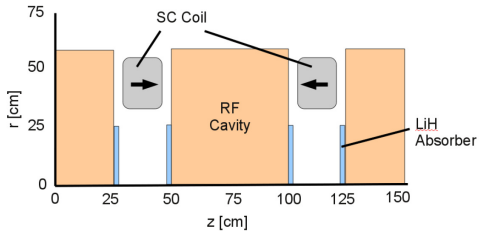

**Figure 3:** Radial projection of the baseline cooling channel cell layout, with LiH absorbers, cylindrical rf cavities, and focusing coils.

## REOPTIMIZATION FOR A $\mu^+$-$\mu^-$ COLLIDER

The neutrino factory requires bunch trains that maximize the total number of µ's; the number of bunches is unimportant. However, for maximum luminosity the collider requires $\mu^+$ and $\mu^-$ to be recombined into a small number of bunches and to be cooled to small longitudinal and transverse emittances. To facilitate recombination the initial capture should be in a shorter bunch train. This can be obtained by capturing at higher momentum within a shorter system.

An evolutionary variant of the IDS baseline with this purpose was developed. In that variant, reference particle 0 was moved to 280 MeV/c, and the Drift, Buncher and Rotator lengths were reduced to ~55m, 31.5 and 36m, respectively, with rf gradients increased to 15, 16 and 18 MV/m, respectively. The focusing fields were increased to 2T. The bunch spacing between the 280MeV/c and 154 MeV/c reference is 10 rf wavelengths (N=10) These incremental changes compress the ~80 m bunch train to ~54 m, so that a similar number of muons are obtained within ~16 bunches.

We also considered further iterations in which rf gradients are increased to 15, 18, and 20 MV/m and also in which the spacing between the reference particles was reduced to 8 wavelengths, (N=8) while the transport lengths were further reduced to ~49, 25 and 27m.

Performance estimates and optimizations are obtained using ICOOL[4] and G4beamline[5]. (see fig.5) The N=10 and N=8 examples obtain higher energy bunch trains, with somewhat more muons in the initial bunches of the trains.

These bunches have a greater momentum width which will require cooling. The IDS bunches had a central momentum of ~250MeV/c with a full width of ~±60MeV/c, while the newer ones have $P_\mu$ =~300 ±~90MeV/c.

Overall muon capture is significantly better in this shorter system, partially from the use of higher gradients. The larger central momentum with the larger momentum spread acceptance captures muons over a broader initial production momentum range. The newer system accepts muons produced from π's with momentum from ~100 MeV/c to ~600 MeV/c, while the IDS accepted from ~100 to 400MeV/c. The larger acceptance matches the production momentum spread more closely.

Table 1 presents some simulations results on muon capture obtained with ICOOL simulations from initial π distributions generated from 8GeV protons on a Hg-jet target (using MARS[6]). The simulations show relatively more $\mu^-$ than $\mu^+$ captured in each case; probably from higher initial production (simulated) in the acceptance. Limiting acceptance to the best 20, 16, and 13 bunches for the three cases captures 2/3 of the total produced muons, although there are extended trains of low-density bunches.

In these initial scenarios we have retained the transverse cooling system developed for the neutrino factory, but with thicker absorbers and higher rf gradients. The cooling system reduces transverse emittances from ~0.02 to ~0.007 m. For a collider we need to transition to cooling systems that include longitudinal cooling, which require dispersive transport with wedge-tyoe absorbers. In future simulations we need to add an initial longitudinal cooling section and explore matching optimizations.

As in the neutrino factory, we chose 201.25 MHz as our baseline frequency. A lower-frequency baseline would make it easier to obtain a smaller number of initial bunches. A higher-frequency case may enable higher gradients, but it would become difficult to fit the large initial beam from the target into smaller higher-frequency rf cavites.

The Front End begins with 4MW of protons on target and ends with ~50kW of muons. A large number of secondaries are produced and are transported down the channel with losses throughout that channel. Much of that beam power is deposited on the walls of the beam transport, and beam losses are ~100W/m in much of the channel, and these losses may overheat components and inhibit hands-on maintenance. This effect must be studied and procedures to limit and manage the losses must be developed.

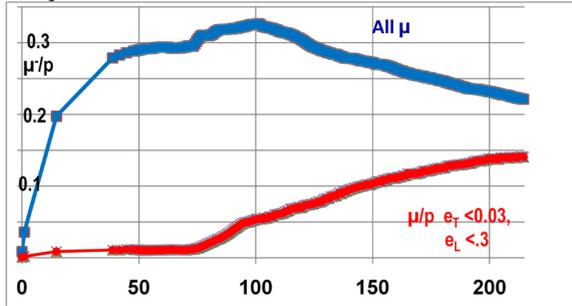

**Fig. 4:** Total μ⁻ and μ⁻ within the acceptance criteria along the N=8 Front end. Transverse cooling after z=100m increases μ's within the acceptance.

## VARIATIONS & FUTURE STUDIES

We have presented approaches to extending the Neutrino factory front end design to the more demanding requirements of a $\mu^+$-$\mu^-$ collider. Variations that improve performance and/or reduce cost will be considered and developed. We will also consider variants that are dramatically different from this evolutionary approach.

## REFERENCES

[1] "International Design Study for the Neutrino Factory", IDS-NF-20, January 2011.
[2] M. Appollonio et al., "Accelerator Concept for Future Neutrino Facilities", RAL-TR-2007-23, **JINST 4** P07001 (2009).
[3] "Cost-effective Design for a Neutrino Factory", with J. S. Berg *et al.*, **Phys. Rev. STAB 9**, 011001(2006).
[4] R. Fernow, "ICOOL", Proc. 1999 PAC, New York, p. 3020, see http://pubweb.bnl.gov/people/fernow/
[5] T. Roberts, G4beamline, ,http://g4beamline.muonsinc.com.
[6] N. V. Mokhov et al., http://www-ap.fnal.gov/MARS/. FERMILAB-Conf-04/053-AD, Proc. 10th Int. Conf. on Radiation Shielding, Funchal, Portugal (2004).

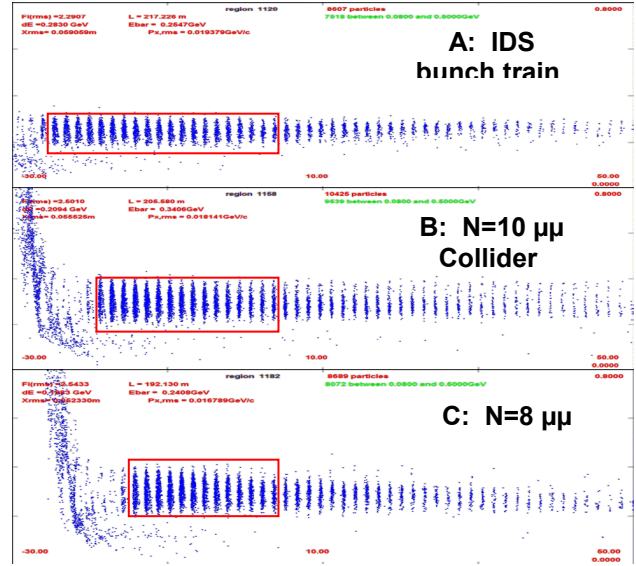

**Fig. 5.** Longitudinal projections of the μ⁻ bunch trains developed in Front End simulations. A: IDS neutrino factory front end at z=217m. B: N=10 μ+-μ⁻ Collider front end at z=207m. C: N=8 example at z=192 m. In each plot the vertical axis is momentum (0 to 0.8 GeV/c) and the horizontal axis is longitudinal position (cτ) (-30 to 50m). Bunches to be recombined are outlined in red.

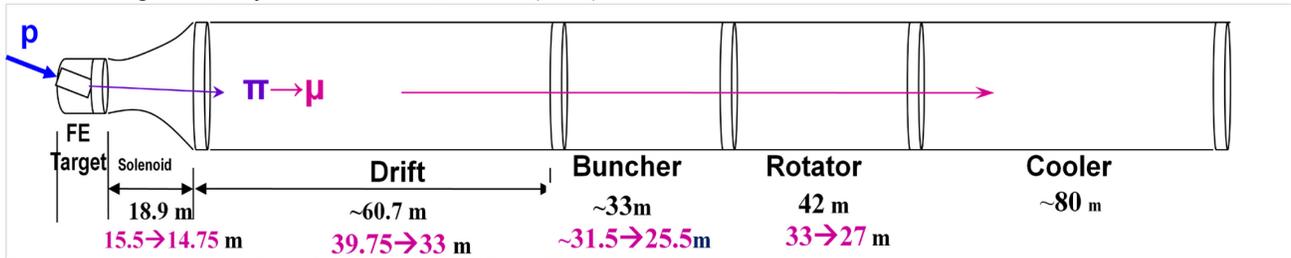

**Figure 1:** Overview of the IDS neutrino factory front end, consisting of a target solenoid (20 T), a tapered capture solenoid (20 T to 1.5T, 19m long), Drift section (~60m), rf Buncher (33 m), an energy-phase Rotator (42m), and a Cooler (~80m). In developing the N=10 and N=8 variants more suited to a $\mu^+$-$\mu^-$ Collider, we reduce the section lengths (as indicated in magenta in the figure) while increasing rf and focusing fields.

**Table 1**: Comparison of muon source front end systems.

| Front end Scenario | Drift, Buncher, Rotator Length | Rf Voltages (Buncher, Rotator, Cooler) | Full length w/ 75m cooling | μ⁺/p ($\varepsilon_t$<0.03, $\varepsilon_L$<0.3m) | μ⁻/p ($\varepsilon_t$<0.03, $\varepsilon_L$<0.3m) | Core bunches, $N_B$, all μ⁻/p |
|---|---|---|---|---|---|---|
| IDS/NF | 80.6, 33, 42m | 0→9, 12, 15 | 230m | 0.086 | 0.116 | 20/0.107 |
| N=10 | 55.3, 31.5, 33 | 0→12, 15, 18 | 205 | 0.106 | 0.143 | 16/0.141 |
| N=8 | 47.8, 25.5, 27 m | 0→15, 18, 20 | 180 | 0.102 | 0.136 | 13/0.123 |